\documentstyle[sprocl,epsf]{article}

\def\be{\begin{equation}}
\def\ee{\end{equation}}
\def\bea{\begin{eqnarray}}
\def\eea{\end{eqnarray}}

\begin{document}

\title{
FRAGMENTS IN GAUSSIAN WAVE-PACKET DYNAMICS\\
WITH AND WITHOUT CORRELATIONS\footnote{Talk given at the 17th
Int.\ Symposium on Innovative Computational Methods in Nuclear
Many-Body Problem, Osaka, Japan, November 10-15, 1997.}}

\author{P.~DANIELEWICZ$^{b, \dag}$, D.~KIDERLEN$^{b,c,\ddag}$\\[1ex]}

\address{$^{b}$National Superconducting Cyclotron Laboratory and\\
Department of Physics and Astronomy, Michigan State University,
\\
East Lansing, Michigan 48824, USA\\[2ex]
$^c$Physik-Department, TU M\"unchen\\
D-85747 Garching, Germany\\[2ex]
E-mail: $^\dag$danielewicz@nscl.msu.edu\\
\hspace*{6em} $^\ddag$kiderlen@physik.tu-muenchen.de
}

\maketitle\abstracts{
Generalization of Gaussian trial wave functions in quantum
molecular dynamics models is introduced, which allows for
long-range correlations characteristic for composite nuclear
fragments. We~demonstrate a~significant improvement in the
description of light fragments with the correlations.
Utilizing
either type of Gaussian wave functions, with or without
correlations, however, we~find that we cannot describe
fragment formation in a~dynamic situation.  Composite fragments
are only produced in simulations if these fragments are present
as clusters in the substructure of original nuclei.
The~difficulty is traced to the delocalization of wave
functions during emission.  Composite fragments are produced
abundantly in the Gaussian molecular dynamics in the limit
$\hbar \rightarrow 0$.
}

\section{Correlated Molecular Dynamics}
This talk is devoted to the description of fragments and their
production in models of molecular dynamics with and without
correlations.  The~considered correlations are
long-range, such as characteristic for composite
fragments,
different from the short-range correlations discussed by
H.~Feldmeier.  In the talk I~shall, in sequence, discuss the
introduction of correlations into molecular dynamics,
describe the corresponding equations of motion, and discuss
results for light clusters: deuteron and alpha.
Subsequently, I shall turn to reactions and discuss results
from simulating the decay of an excited system formed in
reactions and
results from collision calculations.  I~will end the talk with
conclusions.

In quantum molecular dynamics models, the~exact
$N$-particle
wave function of a~system is replaced by a~trial wave function
used in the time-dependent variational principle.
The~different molecular approaches include Quantum Molecular
Dynamics
(QMD)~\cite{aic86,mar90}, Antisymmetrized Molecular Dynamics
(AMD)~\cite{ono92}, Extended Quantum Molecular Dynamics
(EQMD)~\cite{mar96}, and Fermion Molecular Dynamics
(FMD)~\cite{fel90,fel95}.  A~common feature of these approaches
is that the trial wave function is a~product, either
antisymmetrized
or not, of the single-particle Gaussian wave functions
\be
\psi_\nu (\vec{x}) = n \, \exp{\left(-a_\nu (\vec{x} -
\vec{r}_\nu)^2 + i \, \vec{p}_\nu \, \vec{x}\right)} .
\label{1wf}
\ee
In effect of the product assumption, correlations between
nucleons, other than those
possibly induced by the antisymmetrization, vanish.
The~molecular approaches, otherwise, differ in some important
details.
Thus, QMD is the most classical of the approaches;  the~trial
wave function is primarily used in the derivation.  The~width in
the wave function (\ref{1wf}) just acts to
increase the
range of effective interaction; the~dynamics of the width
is suppressed.  No antisymmetrization is carried out and its
effects are simulated using a~Pauli potential.   Within AMD
the antisymmetrization is carried out.  In~EQMD the spreading
of wave functions with time is taken into account.  In~FMD
both the antisymmetrization is carried out and the width
dynamics is included.

In any of the existing approaches, the~center-of-mass
wave function for any nucleon cluster is localized since,
for a~product wave function, the~center of mass gets necessarily
localized once the internal
coordinates are localized.  When a~low-energy nucleon is
emitted
from a~nucleus, it becomes delocalized getting rid of its
localization energy.  That has been, in fact, observed in the
FMD
simulations of reactions~\cite{fel95}.  The~delocalization,
however, cannot
take place in the molecular dynamics models for clusters,
since their intrinsic state must stay localized.
For
light clusters such as deuterons or alphas, the~cm localization
energy is of the order of $\Delta E \sim (10-20)$~MeV.
Given that temperatures are low when the fragments are
expected
to be produced in the low-energy reactions, $T < 10$~MeV,
the~cm localization energy can result in significant
suppression factors for the fragment emission,
${\rm e}^{-\Delta
E/T} < (1/10 - 1/3)$.

Aiming at the description of fragment production in
reactions, we considered ways of solving the problem of the
artificial cm localization caused by the assumptions within the
molecular dynamics.  Let us first take a~look at a~deuteron
Gaussian
wave function, in which the intrinsic and cm motions are
separated,
\bea
\nonumber
\hspace{-2em}\exp{\Big( - \left( {x_1 + x_2 \over \lambda' }
\right)^2 \Big)} \,
\exp{\Big( - \left( {x_1 - x_2 \over \lambda } \right)^2
\Big)}  = \hspace*{3em}\\ \hspace{-1em}
\exp{\Big( - (x_1 \, x_2) \left( \begin{array}{cc}
{1 \over \lambda'^{2}} + {1 \over \lambda^{2}} \hspace*{1em} &
{1 \over \lambda'^{2}} - {1 \over \lambda^{2} }
\\[1ex] {1 \over \lambda'^{2}} - {1 \over \lambda^{2}}
\hspace*{1em}  &
{1 \over \lambda'^{2}} + {1 \over \lambda^{2}} \end{array}
\right) \left( \begin{array}{c} x_1 \\[1ex] x_2 \end{array}
\right) \Big)} \, .
\label{dwf}
\eea
The wave function that is a~product of Gaussians in the cm and
relative coordinates represents an exponential of the bilinear
form in the coordinates of individual nucleons in~(\ref{dwf}).
The~case
of a~delocalized deuteron with total momentum tending to zero
corresponds to $\lambda'
\rightarrow \infty$ in (\ref{dwf}) and a~small
finite~$\lambda$.  The~two nucleons are then correlated: the cm
position may be anywhere in space but if one nucleon is
found somewhere, then the~other nucleon is within the distance
of the order
of $\lambda$ away from the first.  On~the
other hand, the~wave function reduces to a product of
single-particle wave functions only when $\lambda = \lambda'$.

Equation~(\ref{dwf}) suggests to use a generalized Gaussian as
a~trial wave function for $N$ particles,
\be
\langle  x_1,\dots, x_N |\Psi \rangle =
 {\cal N} \, \exp{\left[-A_{ij}\,(\vec{x}_i
-\vec{r}_i) \cdot (\vec{x}_j-\vec{r}_j) +i \, \vec{p}_i
\cdot \vec{x}_i \right]} \, ,
\label{gG}
\ee
where $A$ is the width matrix.
The generalization (\ref{gG}) of the product of Gaussian
single-particle wave functions allows for a~full separation of
the cm and intrinsic coordinates for any nucleon cluster and
introduces correlations between different nucleons, $i \ne j$.

The width matrix $A$ in (\ref{gG}) is a~symmetric, complex, $N
\times N$ matrix with a~positive-definite real part.
The parameters $\lbrace \vec{r}_i \rbrace$ and $\lbrace
\vec{p}_i \rbrace $, respectively, represent centroids and
momenta for wave packets and the set
$\lbrace A_{ij},
\vec{r}_i, \vec{r}_i \rbrace$ is related to various
expectation values, including those that quantify the
correlations, by
\bea
\hspace{-1em}\langle \vec{x}_i \rangle & = & \vec{r}_i
 \\\hspace{-1em} \langle -i \, \vec{\nabla}_i \rangle & = &
\vec{p}_i
\\ \hspace{-1em} \langle (\vec{x}_i)_a \, (\vec{x}_j)_b \rangle
& = &
(\vec{r}_i)_a \, (\vec{r}_j)_b +  { \delta_{ab} \over 4} \,
({\rm Re}A)^{-1}_{ij}  \\ \hspace{-1em}
\langle i(\vec{\nabla}_i)_a \, i(\vec{\nabla}_j)_b \rangle
& = & (\vec{p}_i)_a \, (\vec{p}_j)_b +
\delta_{ab} \, (A^\ast \, ({\rm Re}\,A)^{-1} \, A)_{ij} \, ,
\eea
where $a$ and $b$ are indices for cartesian coordinates.  For
standard molecular dynamics, the~width matrix is diagonal in
particle indices
\be
A_{ij} = A_i \, \delta_{ij} \, .
\ee

\section{Equations of Motion}

Equations of motion for the parameters $\lbrace
A_{ij},
\vec{r}_i, \vec{p}_i \rbrace \equiv \lbrace q_\mu \rbrace
 $  follow from the variational principle
\be
\delta \int_{t_1}^{t_2} dt \, \langle\Psi\vert i \, {d\over
dt} -H\vert\Psi \rangle = 0 \, .
\ee
(For unrestricted variation, the principle yields the exact
Schr\"odinger equation.)
The~equations take a~form
\be
{\cal A}_{\nu\mu} \, \dot q_\mu = -{\partial\over\partial
q_\nu} \, \langle H \rangle \, ,
\ee
where
the skew-symmetric matrix ${\cal A}$ represents the product of
wave function derivatives
\be
{\cal A}_{\nu\mu}= 2\, \, {\rm Im}\left\langle
{\partial\over\partial
q_\nu} \Psi \, \Bigg\vert \, {\partial\over\partial q_\mu}
\Psi\right\rangle \, .
\ee

To solve the equations of motion, the~matrix ${\cal A}$ needs
to be inverted.  When no antisymmetrization is carried out,
the~structure of the matrix is
\be
{\cal A} = \left( \begin{array}{cccc} 0 & {\cal B} && \\
- {\cal B} & 0 && \\ & & 0 & I \\ & & -I & 0 \end{array}
\right) \, ,
\ee
where the rows and columns refer, in sequence, to the parameter
subspaces $\lbrace {\rm Re} \, A \rbrace$,  $\lbrace {\rm Im}
\, A \rbrace$, $\lbrace r \rbrace$, and $\lbrace p \rbrace$,
$I$ is the identity in $N$-particle space, and
${\cal
B}$ is a~quadratic matrix of the dimension $N(N+1)/2$.
The~matrix ${\cal B}$ can
be inverted analytically to yield a~result in terms of the
matrix
elements of~$A$.  Because of this analytic inversion, the~final
effort in solving the equations of motion scales with particle
number like $\sim N^4$; with the
antisymmetrization the effort scales like $\sim N^6$.

Explicitly, the equations of motion take the form
\be
\dot{\vec{r}}_i = {\partial \langle H \rangle
\over\partial\vec{p}_i} \,
,\qquad \, \dot{\vec{p}}_i = -{\partial \langle H
\rangle\over\partial\vec{r}_i} \, ,
\label{eomr}
\ee
\be
\hspace{-1em} {d\over dt} \, {\rm Re}\,A =
{\cal B}^{-1} \,
{\partial \langle H \rangle \over\partial \, {\rm Im}\,A}
\, ,\qquad \,
{d\over dt} \, {\rm Im}\,A=-{\cal B}^{-1} \, {\partial
\langle H \rangle \over\partial \, {\rm Re}\,A}  \, . \\[.5ex]
\label{eomA}
\ee
Contribution to the rhs of the equations for $A$ from the {\em
kinetic energy} can be written as
\be
\hspace{-1em} {\cal B}^{-1} \, {\partial\langle T\rangle\over\partial \,
{\rm Im}\, A}= {2 \over m} \, {\rm Im}\, A^2, \qquad
{\cal B}^{-1} \, {\partial\langle T\rangle\over\partial \, {\rm
Re}\, A}= {2 \over m} \, {\rm Re}\, A^2 \, . \\[.5ex]
\label{eomT}
\ee

From (\ref{eomA}) and (\ref{eomT}), the~$A$-matrix equation for
free particles is
\be
\dot A= - {2 i \over m} \, A \, A
\ee
with a solution
\be
A(t) = A_0 \, \left( 1+ {2 i \over m} \, (t-t_0)
\,A_0 \right)^{-1} \, .
\label{At}
\ee
For initial conditions in a~Gaussian form,
the~solution~(\ref{At})
represents an~exact solution to the Schr\"odinger equation,
with
the spreading in time in the relative and cm coordinates for
any set of
nucleons.  If the matrix~$A$ is initially diagonal in these
coordinates within the $N$-particle space, it will
stay diagonal in these coordinates
later on. When an~interaction is included on the rhs of
equations
(\ref{eomA}), it~affects the evolution of the matrix elements
and spreading for the intrinsic motion, but not for the cm
motion.  When, though, the~matrix~$A$ is always restricted to
a~diagonal form in the single-particle coordinates,
then the~intrinsic
and cm motions get coupled.  If~spreading in the intrinsic
coordinates is limited, so becomes the spreading in the cm
coordinates.  The intrinsic and cm
energies are not separately conserved.  Consequences of the
coupling will be illustrated for light clusters.

\section{Light Cluster Description}

\begin{figure}[t]
\begin{center}
\epsfysize=3.8in \epsffile{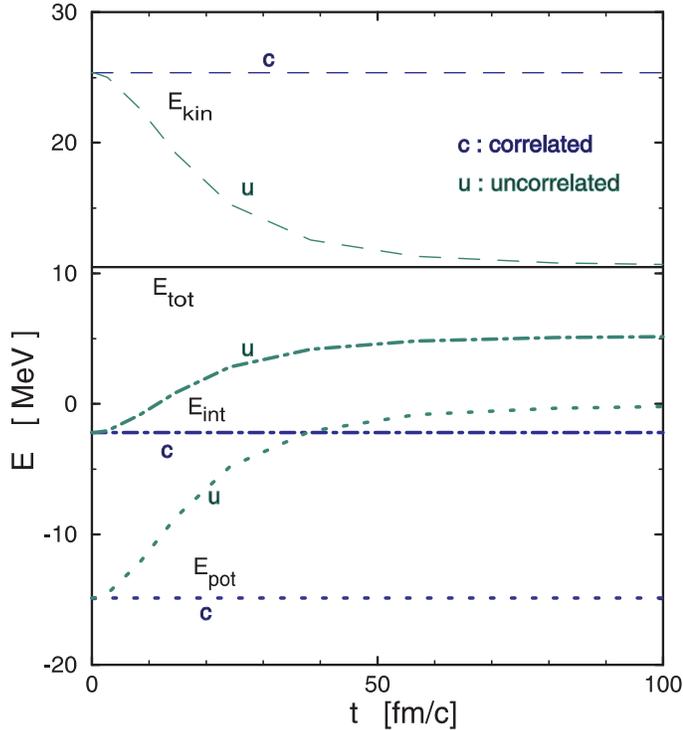}
\end{center}
\caption{
The~dashed, solid, dash-dotted, and dotted lines show,
respectively, the~kinetic, total,
internal, and potential energies, as a~function of time, for
a~free deuteron
in correlated~($c$) and and uncorrelated~($u$) dynamics.
  }
\label{deuteronE}
\end{figure}

\begin{figure}[t]
\begin{center}
\epsfysize=3.8in \epsffile{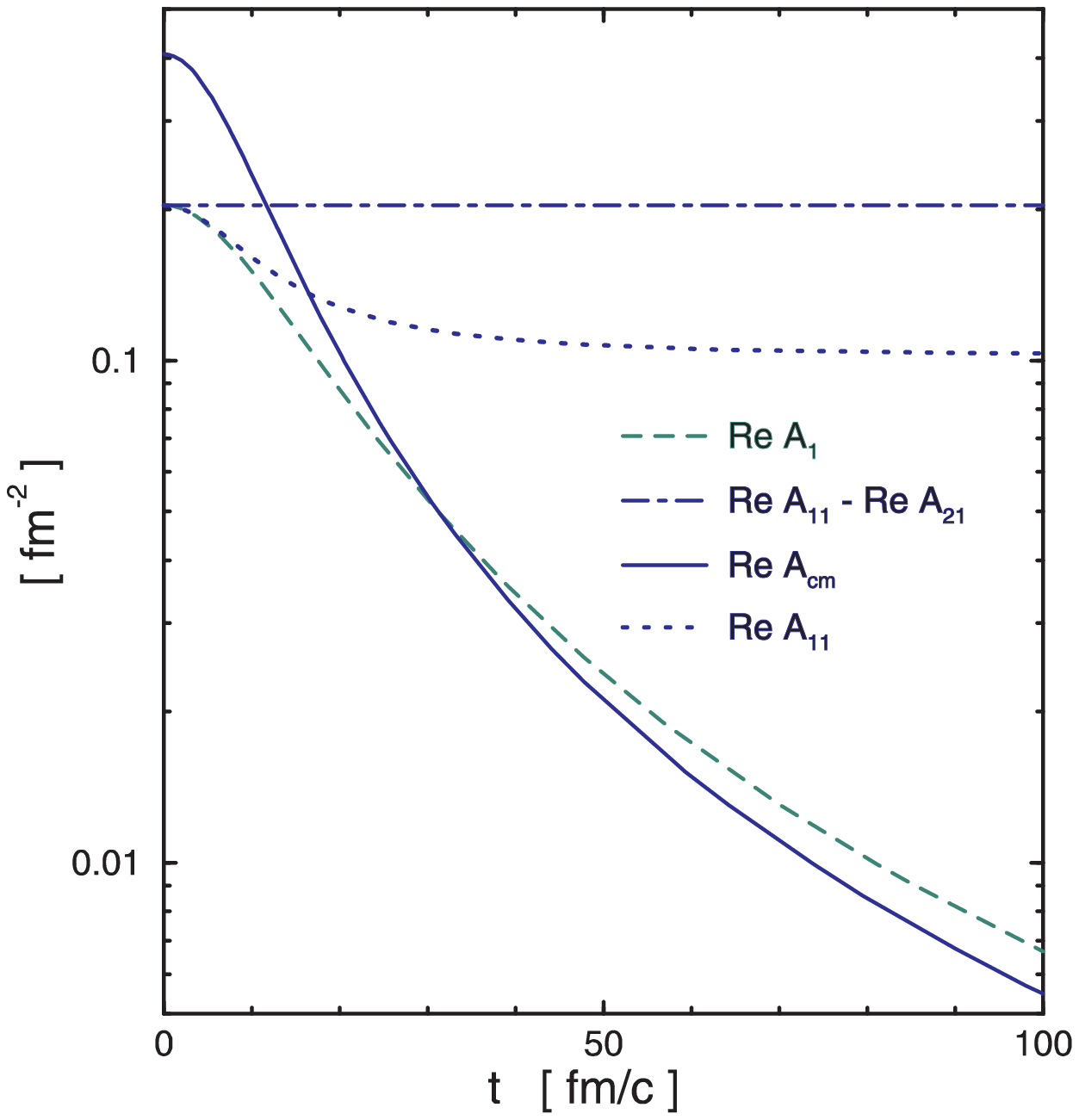}
\end{center}
\caption{
Evolution of the elements of the width matrix for the internal
ground state of a~free deuteron.  The~dashed
lines shows the evolution of the element in the case of
an~uncorrelated
wave function.  The~dash-dotted, solid, and dotted lines show,
respectively, the~evolution of the internal, center-of-mass,
and diagonal elements in the case of a correlated wave
function.
 }
\label{deuteronA}
\end{figure}

Figure \ref{deuteronE}
shows the evolution of different
energies for an~isolated deuteron,
within the standard uncorrelated
wave-packet dynamics ($u$) and within the correlated
dynamics~($c$)
with nonvanishing off-diagonal matrix elements in~$A$,
following Eqs.~(\ref{eomr}) and~(\ref{eomA}).  In~the
correlated dynamics the various energies for the deuteron stay
constant.  In~the uncorrelated dynamics, the~intrinsic motion
taps on the energy from the cm motion.  The~intrinsic energy
increases with time and at $t \sim 12$~fm/c the deuteron
becomes unbound.  Figure~\ref{deuteronA}
shows the evolution of
the elements of the width matrix for the deuteron in the
uncorrelated (dashed
line) and correlated (remaining lines) dynamics.  In the
correlated dynamics the intrinsic width (corresponding to the
dot-dashed line) stays constant, while the cm width (solid
line) behaves like one for a~free particle with twice the
nucleon mass.  The~single-particle element $A_{11}$ (dotted
line) drops with time and then stabilizes.  In~uncorrelated
dynamics
the single-particle element drops continuously; the~deuteron
dissolves.

The intrinsic state of an alpha particle initialized in the
lowest state of intrinsic energy is stationary in the
correlated wave-packet dynamics~\cite{kid97}.  The~cm
state evolves in the same manner as the state for a~particle
with four times the nucleon
mass.  In~the uncorrelated dynamics, on the other hand,
the~alpha particle,
initialized as in the correlated dynamics, pulsates with time
due to the energy exchange between the intrinsic and cm degrees
of freedom.  The~energy acquired from the cm is not large
enough in this case to dissolve the particle.

The results for the isolated particles show the benefits of
using the correlated over the uncorrelated wave functions in the
fragment description.  We~now turn to fragment production.

\section{Reactions}

Two types of simulations relevant for heavy-ion reactions will
be discussed.  One will be the decay of an~excited system
formed in the reactions.  Second type will be complete
reaction simulations followed from an~initial state.

In our decay simulations we initialized the system (e.g.\ of
mass $A = 80$)
in the state of a~local thermal equilibrium with radial flow:
The~centroids
$\vec{r}_i$ were randomly selected within a~sphere of
radius $R = (4 \pi \rho/3)^{-1}$ ($R=5$~fm).
The
momenta $\vec{p}_i$ were selected randomly from a~finite-$T$
Fermi-distribution ($T= 5-12$~MeV).
To~every $\vec{p}_i$, a~radial-flow momentum was added,
proportional to position $\vec{p}_i \propto \vec{r}_i$
($E_{col} = 0-25$~MeV/nucleon).
The width matrix was either taken proportional to unity
($\Delta x = 1.9$~fm),
or its diagonal and off-diagonal elements were generated
randomly according to a~distribution, or, finally, the matrix
elements were equilibrated by keeping the system, at first, in
a~narrow external oscillator potential.

\begin{figure}[t]
\begin{center}
\epsfysize=3.8in \epsffile{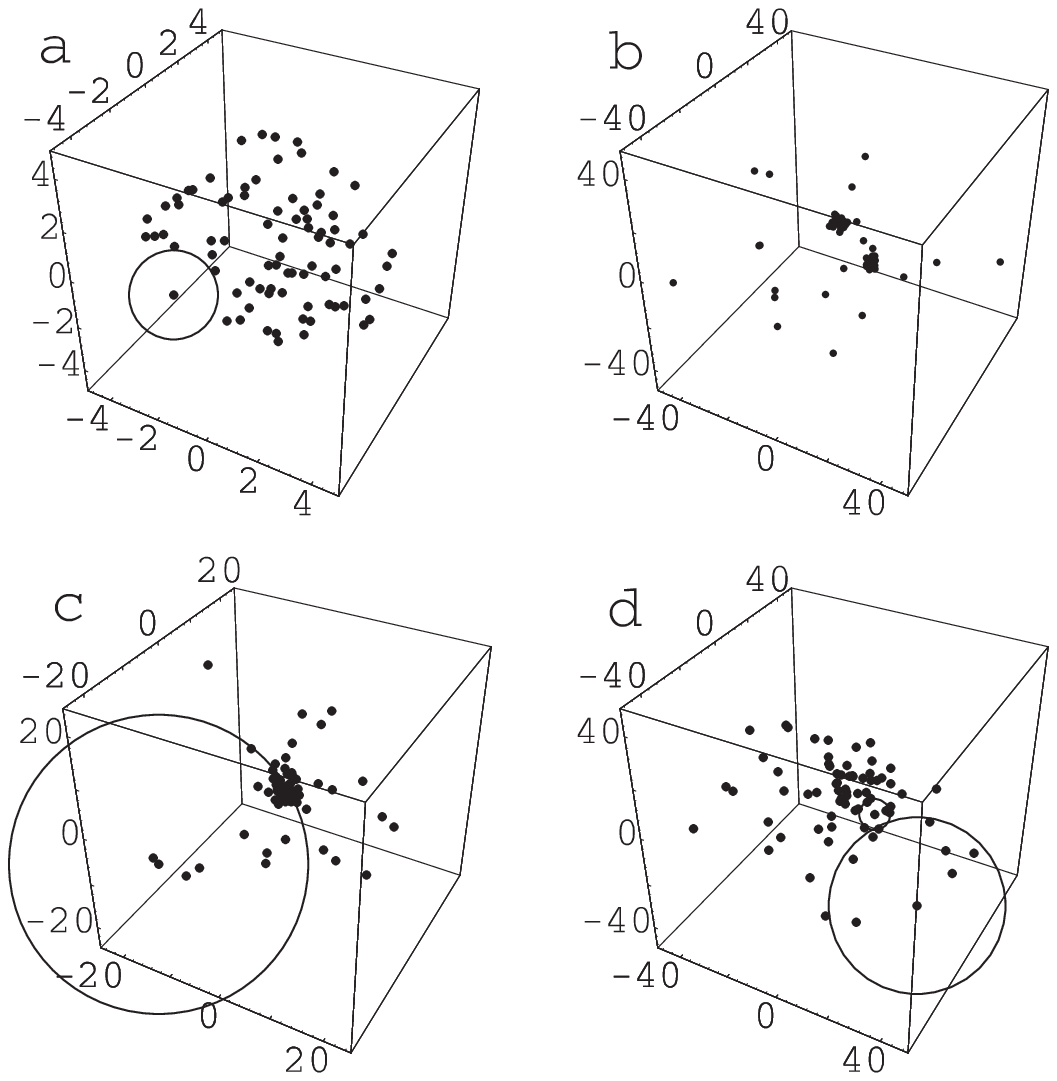}
\end{center}
\caption{
Particle centroids (dots) in the configuration space for
$A=80$, in the initial state~(a) and at $t=200$ {\rm fm/c}
for the static wave-packet width~(b), and for the dynamic
width matrix~(c) and~(d). In~(c), the initial flow energy
is lowered, compared to (b) and~(d), by the initial energy
content in the localization of the wave packets.  The
circles indicate rms radii of the most and least localized
Gaussians.
The~axes show distances in fm.
 }
\label{exc}
\end{figure}

The initial (panel (a)) and late stages (panels (b)-(c)) of
a~system
prepared in the described manner are shown in Fig.~\ref{exc}.
In the calculations with a~dynamic width matrix, the~excited
system emitted a~number of nucleons but no clusters, neither of
intermediate mass (IMF), nor lighter such as deuteron or
$\alpha$ particle.  There are
some centroids in Fig.~\ref{exc}(c) that are close together,
but the associated widths are so large that the interaction
cannot hold the nucleons together and the centroids will
separate with time.

When you get a~final state of simulation
without clusters you think first that it
is a~coincidence and you repeat the simulation with somewhat
changed initial conditions.  Again you find no clusters.  Then
you do that again and we have done that tens of times, maybe
close to a~hundred,
and never saw any fragments, even an~alpha.  We~observed, in
fact little difference in the evolution for centroids between
the uncorrelated dynamics and the correlated dynamics with the
$A$-matrix initialized in a~diagonal form.  The~simulations
initialized with finite off-diagonal elements in~$A$ also
resulted in one residue and many emitted nucleons, but no IMFs
or alphas.

As the primary reason for the lack of fragment production in
the simulations we find the spreading of the wave packets with
time.  By the time the system reaches low densities and the
fragments are expected to form, the~packets get so spread out
that the mutual interaction between the packets is {\em unable
to shrink them} into clusters.  Clearly, in the past, the
production
of IMFs and light clusters has been observed in the QMD
calculations.  Indeed, if we suppress the width dynamics
(corresponding to $\hbar \rightarrow 0$ in our
equations of motion), we begin to observe fragment production.
This is illustrated in Fig.~\ref{exc}(b) that shows a~late
stage of the excited system evolved following the dynamics with
a~frozen wave-packet width.  Two IMFs as well as two deuterons
can be seen in the panel.  Additional small clusters have been
emitted before the shown time and left the displayed spatial
region.

Our results on cluster production for the dynamic width
matrix may seem in contradiction to the FMD
results~\cite{fel95} with even multifragmentation events
reported in nuclear collisions.  Indeed, in~our own
simulations
of collisions~\cite{kid97}, we observed the~production of
clusters, but only if these were present in the substructure of
original nuclei and did not manage to dissolve in the
reaction.  We never observed clusters that were formed in
a~reaction. For
example, in the ground state of $^{12}$C, for our interaction,
the~centroids form three $\alpha$-type clusters of four
nucleons each.  If we collide two $^{12}$C nuclei, one or two
of them may break up into three $\alpha$ particles.  On the
other
hand, if we collide two $^{40}$Ca nuclei, with no internal
substructure, one or two residues
are formed and many nucleons are emitted, but no clusters of
intermediate mass.

\section{Conclusions}

In~general, for uncorrelated wave functions, the~internal state
cannot be localized without localizing the center of mass.
Thermal estimates indicate that the localization can act to
suppress fragment production.  In~the dynamics, the~use of
uncorrelated wave functions leads to unphysical exchange of
energy between the
cm and intrinsic degrees of freedom.  These deficiencies are
eliminated in the dynamics utilizing correlated
wave functions.  The~correlations that we have introduced are of
such nature as expected
for fragments and they allow for the separation of any number
of nucleons in a~wave function.  The~benefits of the correlated
wave functions have been illustrated with the examples of
deuteron and $\alpha$ particle.

Despite the benefits,
in reaction simulations with dynamic widths, with or without
correlations, we observe the cluster production {\em only}
when the clusters are present in the initial state.
The~absence of new clusters is associated with large spreading
of
the wave function at the~time when
the reacting system expands and the new clusters should
form.
While there is nothing unphysical
in the spreading of the wave function, as such, in the
simulations,
in~reality
the~interaction would be capable of pinching portions of two or
more packets and forming fragments out of the portions of these
packets.  The~Gaussian
parametrization of the wave function does not permit for the
process.

On the basis of the reaction simulations,
we conclude that, in a~successful description of fragment
production, the wave function
must have a flexibility
to change over distances comparable to the interaction range,
at a~time when the fragments are formed.
The~widths
of the packets in AMD or QMD may already act to suppress the
fragment production.

\section*{Acknowledgments}
This
work was partially supported by the National Science Foundation
under Grant PHY-9605207.

\end{document}